\documentclass[10pt,twocolumn,letterpaper]{article}

\usepackage{cvpr}              % To produce the CAMERA-READY version
\usepackage{amsmath,graphicx, amsfonts, multirow, tabularx, booktabs, caption, float, placeins}
\usepackage{colortbl}
\usepackage{amssymb}
\usepackage{stmaryrd}
\usepackage[dvipsnames]{xcolor}
\definecolor{nicergreen}{rgb}{0.13, 0.54, 0.13}
\definecolor{nicered}{rgb}{0.83, 0.16, 0.16}
\definecolor{myhighlight}{rgb}{0.91, 0.95, 0.93}
\usepackage[dvipsnames]{xcolor}

\definecolor{cvprblue}{rgb}{0.21,0.49,0.74}
\usepackage[pagebackref,breaklinks,colorlinks,citecolor=cvprblue]{hyperref}
\title{Audio Mamba: Bidirectional State Space Model for Audio Representation Learning}

\author{
    Mehmet Hamza Erol$^{*}$,~~Arda Senocak$^{*}$,~~Jiu Feng,~~Joon Son Chung\\
    Korea Advanced Institute of Science and Technology\\
    {\small $*$~Equal Contribution}
}

\begin{document}
\maketitle
\begin{abstract}
\vspace{-1mm}
Transformers have rapidly become the preferred choice for audio classification, surpassing methods based on CNNs. However, Audio Spectrogram Transformers (ASTs) exhibit quadratic scaling due to self-attention. The removal of this quadratic self-attention cost presents an appealing direction. Recently, state space models (SSMs), such as Mamba, have demonstrated potential in language and vision tasks in this regard. In this study, we explore whether reliance on self-attention is necessary for audio classification tasks. By introducing Audio Mamba (AuM), the first self-attention-free, purely SSM-based model for audio classification, we aim to address this question. We evaluate AuM on various audio datasets - comprising six different benchmarks - where it achieves comparable or better performance compared to well-established AST model. Code is available at \small{\url{https://github.com/mhamzaerol/Audio-Mamba-AuM}}
\end{abstract}    
\vspace{-4mm}
\section{Introduction}
\label{sec:intro}

In recent years, CNNs~\cite{kong2020panns, he2016deep} have been replaced with transformer-based architectures ~\cite{vaswani2017attention, Dosovitskiy2021vit, touvron2021training, gong21b_interspeech} in a paradigm shift in deep learning, as transformers outperform convolutional neural networks. Not only does the performance of transformers exceed that of CNNs, but establishing a unified architecture among many different research fields and tasks — traditionally using completely different models — is another breakthrough~\cite{gong2022uavm, gong2022contrastive, wu2023large, shih2023speechclip, wu2022wav2clip, nagrani2021attention, morgado2020learning, alayrac2020self, akbari2021vatt}. Despite their success, transformers are hindered by their reliance on the computationally intensive self-attention mechanism. The $\mathcal{O}(n^2)$ cost of attention is a natural concern when processing longer sequences. This limitation motivates the exploration of alternative architectures, notably state-space models (SSMs)~\cite{gu2023mamba, gu2021efficiently, smith2022simplified, fu2022hungry} such as Mamba~\cite{gu2023mamba}, which replaces the self-attention mechanism in favor of incorporating time-varying parameters to capture global context efficiently. Recently, the introduction of Mamba~\cite{gu2023mamba} marks a significant advancement in model efficiency for both training and inference, suggesting a potential alternative to transformer-based approaches. Given the universality and scalability of transformers across various tasks, Mamba's potential, coupled with its computational efficiency, is particularly promising for becoming a similarly generic and versatile architecture.

\begin{figure}[t!]
    \centering
    \includegraphics[width=\columnwidth]{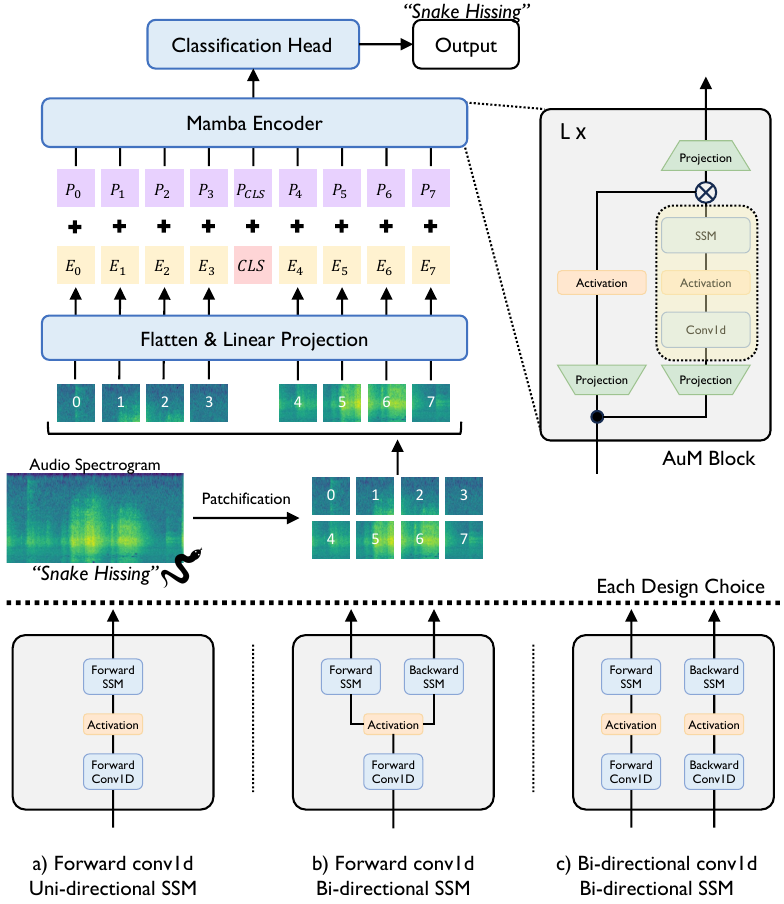}
    \caption{\textbf{The proposed Audio Mamba (AuM) architecture.}}
    \label{fig:pipeline}
\end{figure}

Despite Mamba's recent successes in language modeling and vision~\cite{zhu2024vision, yang2024vivim, liu2024vmamba, ma2024u, xing2024segmamba}, the adoption of Mamba and similar SSM-based models in the audio classification domain still remains unexplored. This gap motivates our work, where we introduce a novel SSM-based model, Audio Mamba - AuM, applied directly to audio spectrograms. Our approach is self-attention free, focusing purely on long sequence modeling with state space models. AuM not only achieves comparable performance to the Audio Spectrogram Transformer (AST)~\cite{gong21b_interspeech}, the most prominent approach in audio classification, but also retains several advantages of transformer-based models. These include the ability to handle varying sequence lengths and the ease of transferability to other tasks. Due to the employment of state space models, reliance on self-attention is eliminated, enabling the model to operate with linear time complexity relative to sequence length and feature dimension, as opposed to AST where quadratic complexity is observed. The closest work to ours is Vision Mamba~\cite{zhu2024vision}, which utilizes bidirectional SSM for global visual context modeling and positional embeddings for location information in a structure similar to Vision Transformers (ViT)~\cite{Dosovitskiy2021vit}. Drawing on AST's success in applying ViT's principles to audio classification, we also draw inspiration from the findings of Vision Mamba and study the methodologies suitable for applying bidirectional state space models to audio classification. To accomplish this task, we take the following steps: (1) We divide the input spectrogram into patches, which are then projected into patch embedding tokens. (2) We add an additional learnable classification token to the sequence of patch tokens, specifically in the middle. (3) The Audio Mamba Encoder blocks process these token sequences in both forward and backward directions with SSM modules. (4) The classification token is utilized to train the model on the supervised audio classification task and also for making predictions in the inference stage. We summarize the contributions of our work as follows:

\begin{itemize}[noitemsep, topsep=0pt]
\item We introduce Audio Mamba (AuM) for processing audio spectrograms, utilizing bidirectional state space models (SSM) to handle tokens in both forward and backward directions, in a similar structure of Audio Spectrogram Transformers (AST).
\item By eliminating self-attention modules, AuM achieves linearly scaled resource consumption when evaluated with long audio sequences.
\item Our comprehensive experiments across six diverse datasets — AudioSet~\cite{gemmeke2017audio}, AudioSet Balanced, VGGSound~\cite{VGGSound}, VoxCeleb~\cite{nagrani2020voxceleb}, Speech Commands V2~\cite{warden2018speech}, and Epic-Sounds~\cite{EPICSOUNDS2023} — show that AuM delivers performance that is comparable to or exceeds the most prominent audio classification method AST.
\end{itemize}

\section{Audio Mamba}
\label{sec:approach}

\subsection{Flow of the Architecture}
The Audio Mamba (AuM) architecture, as depicted in Fig. \ref{fig:pipeline}, begins by transforming an input audio waveform into an audio spectrogram $X \in \mathbb{R}^{F\times T}$, where $F$ and $T$ represent the frequency and time dimensions, respectively. The spectrogram is partitioned into a sequence of $N$ square patches $S \in \mathbb{R}^{N \times p \times p}$, with $p$ denoting the side length of each patch and $N$ calculated as $N=(F/p)\times(T/p)$. Each individual patch $S_i$ is subsequently flattened into a one-dimensional vector $S_i \in \mathbb{R}^{p^2}$, and through a linear projection, it is embedded into a $D$-dimensional space, yielding $E_i \in \mathbb{R}^{D}$. This process is facilitated by the patch embedding layer. Afterward, a special learnable classification token, denoted as $CLS \in \mathbb{R}^{D}$, is inserted into the middle of the sequence, leading to an augmented embedding sequence $E \in \mathbb{R}^{(N+1)\times D}$. To encode the position of each element within the sequence, learnable positional embeddings $P \in \mathbb{R}^{(N+1)\times D}$ are added, resulting in the token sequence $T \in \mathbb{R}^{(N+1)\times D}$. This token sequence is then processed by the Audio Mamba encoder which consists of $L$ stacked blocks, each of which retains the dimensionality of its input. Thus, the encoder transforms $T$ into an output sequence $T^{\prime} \in \mathbb{R}^{(N+1) \times D}$. The modified representation of the classification token $T^{\prime}_{N/2}$ is then conveyed to the classification head.

\subsection{Architecture Details}
Aiming to establish itself as a generic architecture, AuM shares several similarities with the AST~\cite{gong21b_interspeech}. However, AuM distinguishes itself through distinct components and strategic design decisions that highlight its unique architectural and operational characteristics, to be a self-attention-free model.

\noindent\textbf{Preliminaries.} State space models (SSMs) are linear time-invariant systems that aim to model a continuous system which maps a time dependent $D$ dimensional input sequence $x(t) \in \mathbb{R}$ to an output $y(t) \in \mathbb{R}$ through maintaining a hidden state $h(t) \in \mathbb{R}^{N}$. Such a system could be represented with the following equation:
\begin{align}
    \begin{split}
        h^{\prime}(t) &= Ah(t) +Bx(t),\\
        y(t) &= Ch(t).
    \end{split}
\end{align}
where $A \in \mathbb{R}^{N \times N}$, $B \in \mathbb{R}^{N \times D}$ and $C \in \mathbb{R}^{D \times N}$. With the primary goal of adapting the model to deep learning algorithms, a discretization process is applied, which transforms the continuous parameters $A$ and $B$ through a discretization rule into $\Bar{A}$ and $\Bar{B}$, respectively.
These discretized parameters are then substituted for $A$ and $B$, leading to the following discretized formulation of the system:
\begin{align}
    \begin{split}
        h_t &= \Bar{A}h_{t-1} + \Bar{B}x_t,\\
        y_t &= Ch_t.
    \end{split}
    \label{disc}
\end{align}
Such a linear time-invariant system could be computed both as a linear recurrence or through a global convolution, enabling efficient processing~\cite{gu2023mamba}. Despite its efficiency, such a system has limitations in modeling certain types of data due to its time-invariant parameterization. Mamba upgrades existing works based on such models by converting the time-invariant parameters into a time-variant format, enabling efficient derivation of parameters from time-varying inputs. Specifically, inside the Forward SSM module of a Mamba block (Fig.~\ref{fig:pipeline} (a)), the algorithm utilizes the input sequence $x \in \mathbb{R}^{L \times D}$ that has been convolved through a Forward Conv1D before, to convert each time-invariant parameter $A$, $B$, and $\Delta$ in eq.~\eqref{disc} into specific corresponding ones $A^{\prime}_i$, $B^{\prime}_i$, and $\Delta^{\prime}_i$ for each element $x_i$ of the sequence. Mamba then utilizes these parameters by adopting a modern hardware-oriented scanning method that processes the input sequence from beginning to end in a unidirectional manner. More details can be seen in~\cite{gu2023mamba}. This enables the model to selectively update its hidden state by capturing relevant information from the input sequence through these converted parameters.

\begin{table*}[t!]
\centering
\renewcommand{\tabularxcolumn}[1]{m{#1}}
\begin{tabularx}{\textwidth}{ 
    >{\centering\arraybackslash}l 
    >{\centering\arraybackslash}c  
    >{\centering\arraybackslash}c 
    >{\centering\arraybackslash}c  
    >{\centering\arraybackslash}c  
    >{\centering\arraybackslash}c 
    >{\centering\arraybackslash}c
}
\toprule
\multirow{2}{*}{\textbf{Model}} & \textbf{AudioSet} & \textbf{AS-20K} & \textbf{VGGSound} & \textbf{VoxCeleb} & \textbf{Speech Comm. V2} & \textbf{Epic-Sounds}\\
& \textbf{(mAP)} & \textbf{(mAP)} & \textbf{(Acc.)} & \textbf{(Acc.)} & \textbf{(Acc.)} & \textbf{(Acc.)}\\
\toprule
AST-B/16 & 29.10{\tiny $\pm$ 0.07} & 10.41 {\tiny $\pm$ 0.32} & 37.25 {\tiny $\pm$ 0.31} & 22.44 {\tiny $\pm$ 0.19} & 85.27 {\tiny $\pm$ 1.07} & \textbf{44.76} {\tiny $\pm$ 0.20}\\
\midrule
\rowcolor{lightgray!25}
\textbf{AuM-B/16} & {\textbf{32.43} {\tiny $\pm$ 0.31}} {\tiny \textcolor{nicergreen}{(+3.33)}} & {\textbf{13.28} {\tiny $\pm$ 1.07}} {\tiny \textcolor{nicergreen}{(+2.87)}} & {\textbf{42.58} {\tiny $\pm$ 0.28}} {\tiny \textcolor{nicergreen}{(+5.33)}} & {\textbf{28.34} {\tiny $\pm$ 3.38}} {\tiny \textcolor{nicergreen}{(+5.90)}} & {\textbf{91.58} {\tiny $\pm$ 3.17}} {\tiny \textcolor{nicergreen}{(+6.32)}} & {44.17 {\tiny $\pm$ 0.58}} {\tiny \textcolor{nicered}{(-0.60)}}\\
\bottomrule
\end{tabularx}
\caption{\textbf{Results of from-scratch training of AST and AuM base models across various datasets.}}
\label{tab:baseNOinit}
\end{table*}
\noindent\textbf{Bidirectional Mamba Encoder.}
Even though Mamba's unidirectional scan of the sequence offers promising benefits for modeling causal sequential data, its application to learning from 2D data benefits from processing in multiple directions~\cite{zhu2024vision, yang2024vivim, liu2024vmamba}. For instance, in learning from visual data, an existing Mamba-based architecture, ViM~\cite{zhu2024vision}, modifying the original Mamba block in Fig.~\ref{fig:pipeline} (a) to Fig.~\ref{fig:pipeline} (c) by introducing another direction for feature extraction (Backward Conv1D) or scanning (Backward SSM) of the input sequence, enabling multi-directional and spatial-aware processing. Similarly, AuM adopts the design strategy shown in Fig.~\ref{fig:pipeline} (b) by adding an extra backward scanning direction to the original Mamba block. This approach utilizes the same convolved features while adapting both the forward and backward SSM parameters into their time-variant (input-dependent) versions for scanning. Likewise ViM, this enables AuM to model the global context in a spatial aware manner, mirroring the functionality of self-attention mechanism in transformers for modeling global context.

\begin{figure*}
  \centering
  \begin{minipage}{.32\textwidth}
      \centering
      \includegraphics[width=\linewidth]{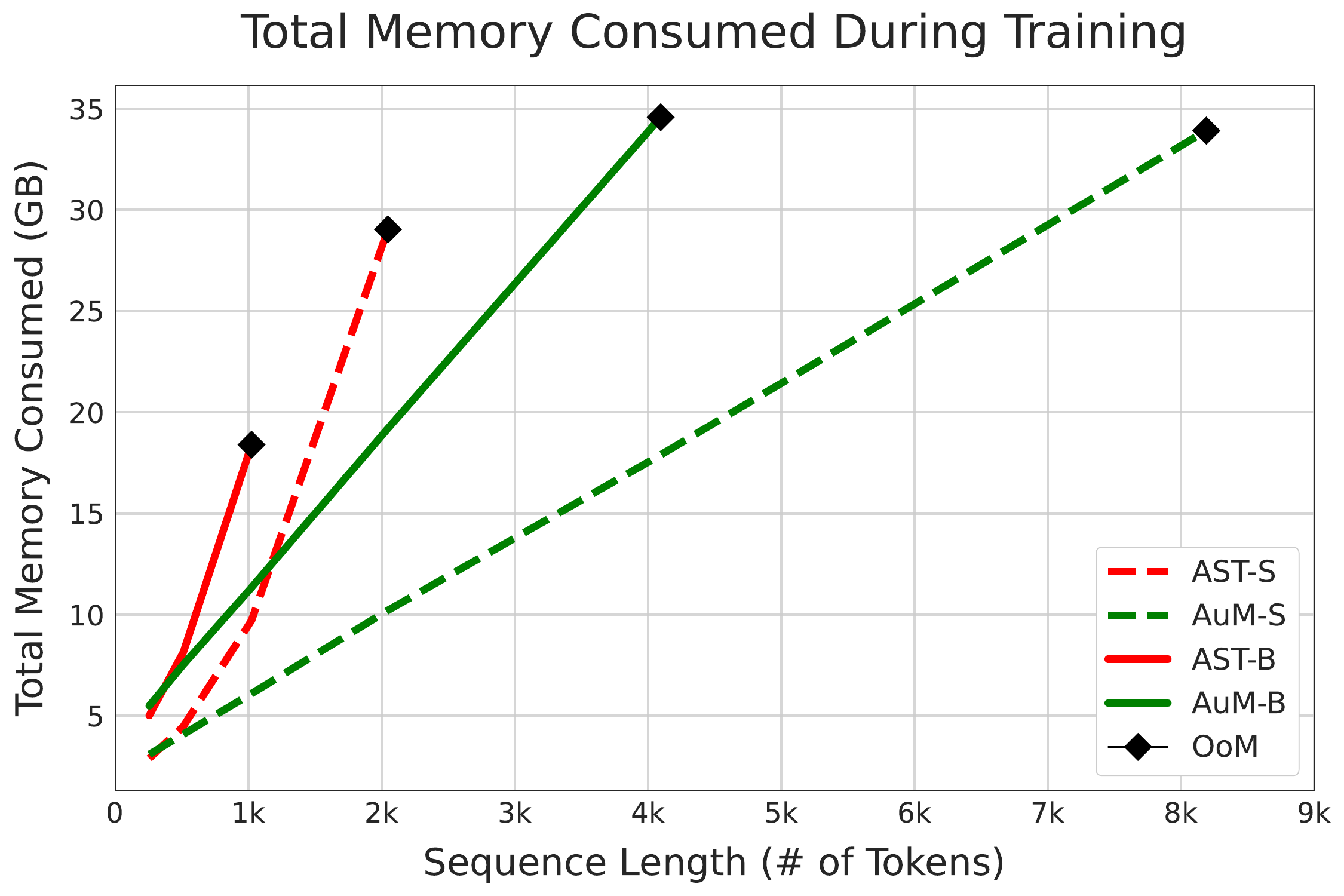}
  \end{minipage}%
  \begin{minipage}{.32\textwidth}
    \centering
    \includegraphics[width=\linewidth]{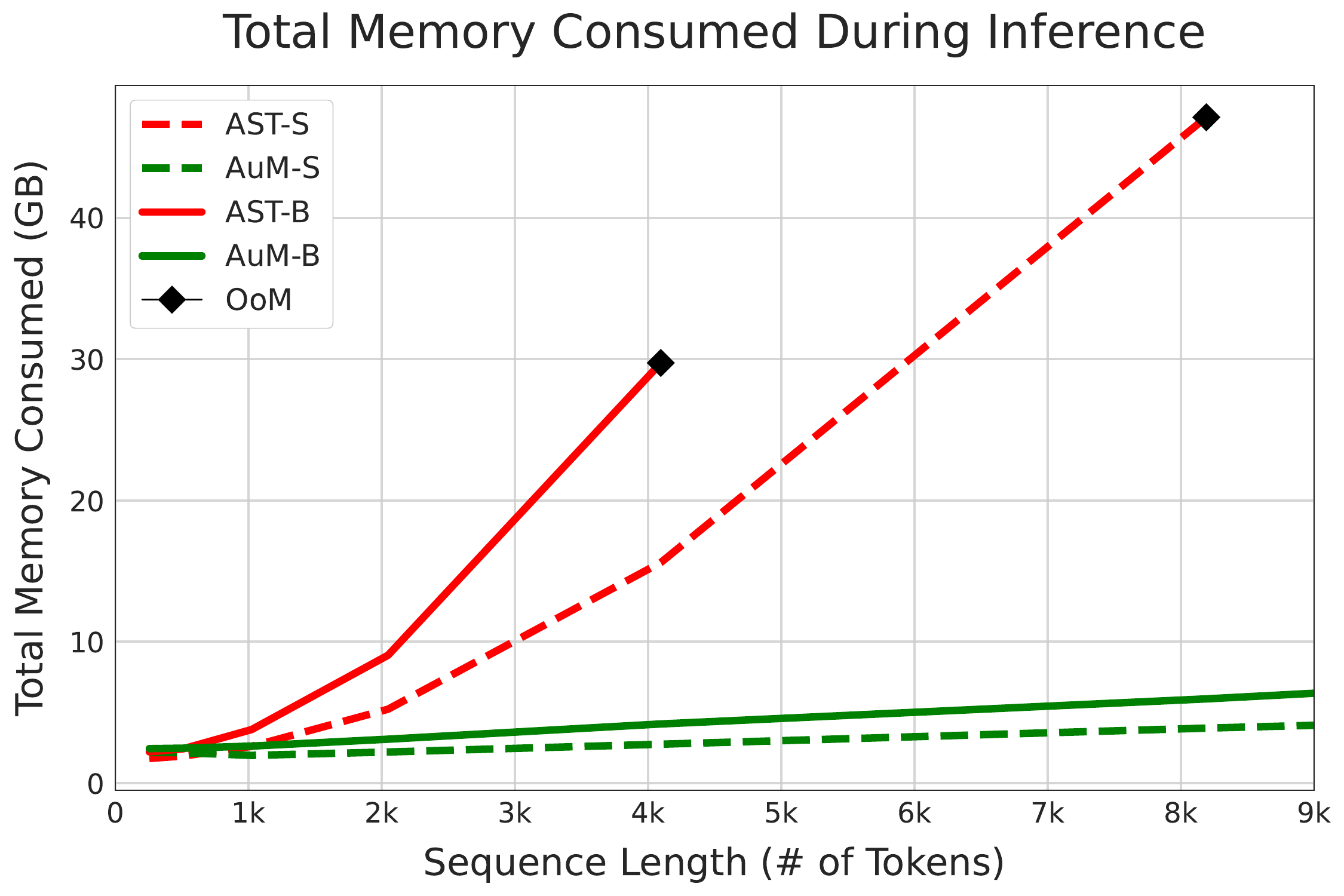}
  \end{minipage}%
  \begin{minipage}{.32\textwidth}
    \centering
    \includegraphics[width=\linewidth]{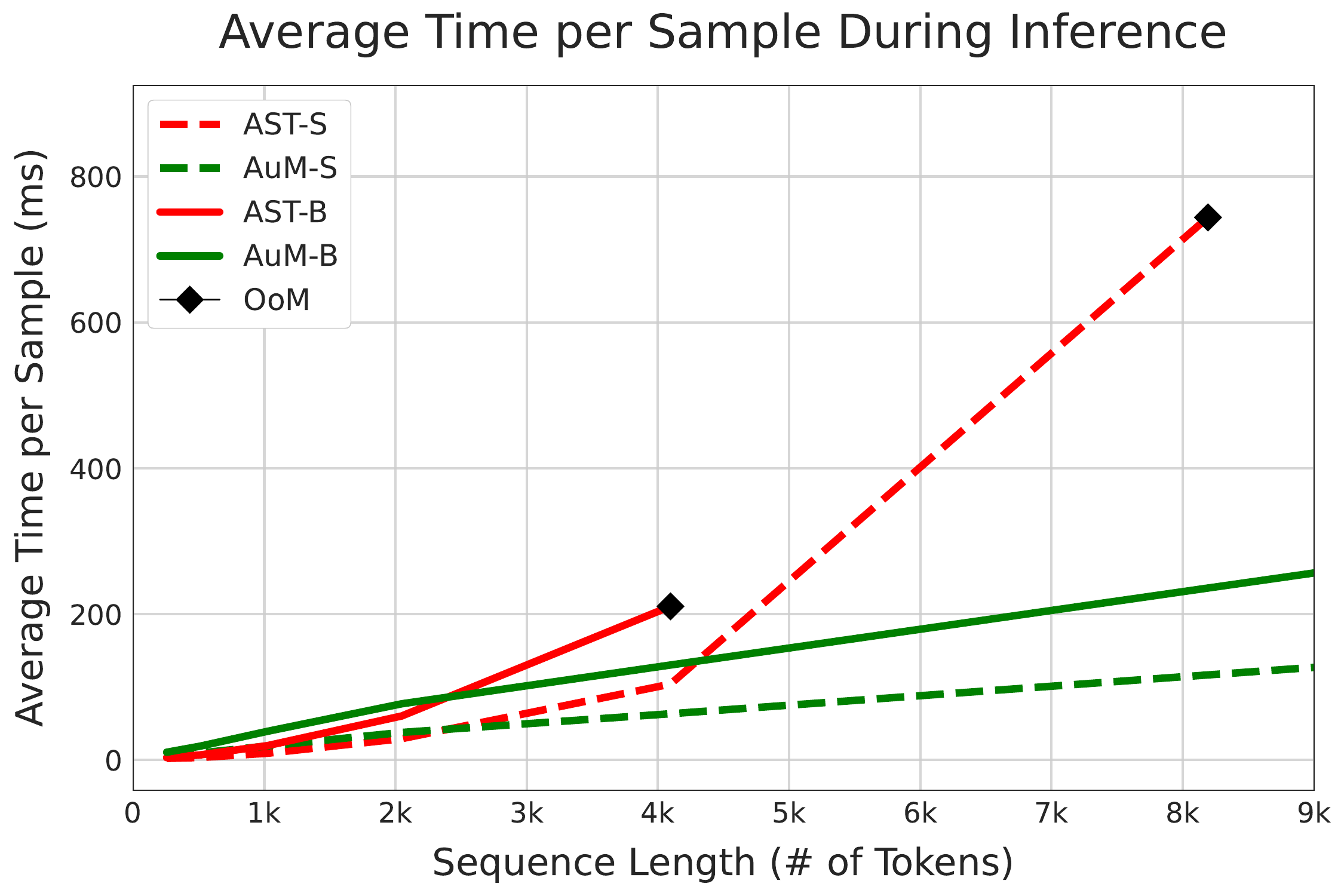}
  \end{minipage}
  \caption{\textbf{Empirical evaluation of memory and time consumption for AST and AuM small/base models.}}
  \label{fig:resource}
\end{figure*}

\noindent\textbf{Classification Token.} Unlike transformers in their pure form, which are permutation invariant when processing the input sequence, AuM block is sensitive to the order of the input sequence because both feature extraction (Conv1D) and SSMs are input-order-sensitive operations. Consequently, in addition to scanning directions, the placement of the classification token within the input sequence becomes critical for the learning process. Similarly to ViM, AuM strategically positions the classification token at the midpoint of the input sequence, immediately after the patch embedding layer. This setup has shown improved performance in bidirectional processing setup, as demonstrated in ViM, and the ablation study conducted in Section~\ref{sec:experiments}. 
\section{Experiments}
\label{sec:experiments}
\subsection{Datasets and Evaluation Metrics}
\textbf{Datasets.} Our experiments utilize: (1) Audioset Full / Balanced, (2) VGGSound, (3) VoxCeleb, (4) Speech Commands-V2, and (5) EPIC-SOUNDS datasets. \textbf{AudioSet}~\cite{gemmeke2017audio} is an expansive dataset with a wide array of audio samples, each marked with a set of labels. It includes over 2 million 10 seconds long audio clips with a total of 527 distinct labels. The balanced set on the other hand is curated from the full set, consisting of 20K samples. \textbf{VGGSound}~\cite{VGGSound} contains nearly 200k video clips of 10 seconds each, annotated with 309 diverse sound categories. \textbf{VoxCeleb}~\cite{nagrani2020voxceleb} is a dataset focused on audio-visual representations of human speech, featuring 1,251 speakers and around 145k speech instances. \textbf{Speech Commands-V2}~\cite{warden2018speech} comprises approximately 105k audio recordings, each with a duration of 1 second, and includes 35 widely recognized speech commands. Finally, \textbf{EPIC-SOUNDS}~\cite{EPICSOUNDS2023}, part of EPIC-KITCHENS-100~\cite{Damen2022RESCALING}, comprises 75.9k audio segments from egocentric videos, labeled across 44 classes, focusing on actions discernible by sound, such as object collisions with material annotations.\\
\textbf{Evaluation metrics.} We utilize mean average precision (mAP) for Audioset experiments due to the existence of multiple labels per sample. For the remaining datasets, we show the top-1 classification accuracy (Acc) as the samples have a single label.

\subsection{Comparison to AST on Standard Benchmarks}
In this section, we conduct a comparative analysis of our Audio Mamba (AuM) against the Audio Spectrogram Transformer (AST) model. Both models use base backbones, AuM-B/16 and AST-B/16. As discussed in the implementation details, we follow the same training and experimental settings as the AST model to ensure a fair comparison, which are detailed at section \ref{sec:train_setup}. It is worth highlighting that in this experiment, neither AuM nor AST utilized pretraining weights from other models (AST is initialized with weights from the Vision Transformer (ViT) model pretrained on ImageNet in the original paper~\cite{gong21b_interspeech}) to ensure a pure comparison of these two different architectures. We repeat each experiment three times with the same setup but different random seeds and report the results with the mean and standard deviation in Table 1. Our proposed \textit{AuM generally achieves better performance} in this experimental setup. This indicates that AuM, with its pure setting, is a potential alternative to the AST model, without relying on self-attention, which leads to better efficiency in computational resources.

\subsection{Comparison to AST on Efficiency}

Transformer-based audio classification models are computationally demanding (quadratic complexity), particularly with lengthy audio and high-dimensional data. SSM-based models stand out for their computational and memory efficiency. In this section, we compare AuM to AST from an efficiency perspective. A single A6000 GPU is used for this experiment. We feed audios with corresponding lengths for every given token number to the models to simulate the speed and GPU memory comparison. We visualize the speed and memory consumption of these models in Figure~\ref{fig:resource}. AuM demonstrates clear computation and memory efficiency. For example, the AuM-Base model that uses 20 seconds of audio (1024 tokens) for training consumes as little GPU memory as the AST-Small model. Additionally, while AuM-B can be trained with 80 seconds, AST-B will run out of memory in the setting that uses only 20 seconds audios. Moreover, AuM is 1.6 times faster in the inference stage than AST at 4096 number of tokens, with a growing rate as the token count increases. All these results indicate that AuM exhibits a trend of linear scaling with respect to sequence length.

\begin{table}[tp]
\centering
\resizebox{1.0\linewidth}{!}{
\scriptsize
\setlength{\tabcolsep}{2.0pt} % Adjust the space between columns
\renewcommand{\arraystretch}{1.2} % Adjust the space between rows
\begin{tabular}{l ccc ccc}
\toprule
\multirow{2}{*}{\textbf{Setting}} & \multicolumn{3}{c}{\textbf{AS-20K (mAP)}} & \multicolumn{3}{c}{\textbf{VGGSound (Acc.)}} \\
\cmidrule(lr){2-4} \cmidrule(lr){5-7}
& \textbf{Head} & \textbf{Mid} & \textbf{End} & \textbf{Head} & \textbf{Mid} & \textbf{End} \\
\midrule
\textbf{Fo-Fo (a)} & 0.48{\tiny $\pm$ 0.00} & 11.73{\tiny $\pm$ 0.47} & 11.90{\tiny $\pm$ 1.05} & 0.33{\tiny $\pm$ 0.00} & 35.05{\tiny $\pm$ 0.80} & 39.09{\tiny $\pm$ 0.56} \\
\rowcolor{lightgray!25}
\textbf{Fo-Bi (b)} & 13.57{\tiny $\pm$ 0.09} & \textbf{13.81}{\tiny $\pm$ 0.32} & 12.35{\tiny $\pm$ 0.33} & 40.91{\tiny $\pm$ 0.47} & \textbf{42.58}{\tiny $\pm$ 0.28} & 40.41{\tiny $\pm$ 0.15} \\
\textbf{Bi-Bi (c)} & 4.97{\tiny $\pm$ 0.13} & 9.69{\tiny $\pm$ 0.43} & 11.11{\tiny $\pm$ 0.66} & 34.22{\tiny $\pm$ 0.26} & 36.48{\tiny $\pm$ 0.46} & 41.09{\tiny $\pm$ 0.16} \\
\bottomrule
\end{tabular}
}
\caption{\textbf{Results of ablation on the design choices.} Architectural choices (under the settings column) refer to the block types in Fig. \ref{fig:pipeline}, the location of the classification token is indicated through the columns: Head, Mid and End per dataset.}
\label{tab:ablation}
\end{table}

\begin{table*}[t!]
\centering
\renewcommand{\tabularxcolumn}[1]{m{#1}}
\begin{tabularx}{\textwidth}{ 
    >{\centering\arraybackslash}l
    >{\centering\arraybackslash}c  
    >{\centering\arraybackslash}c 
    >{\centering\arraybackslash}c  
    >{\centering\arraybackslash}c  
    >{\centering\arraybackslash}c 
    >{\centering\arraybackslash}c
}
\toprule
\multirow{2}{*}{\textbf{Model}} & \textbf{AudioSet} & \textbf{AS-20K} & \textbf{VGGSound} & \textbf{VoxCeleb} & \textbf{Speech Comm. V2 } & \textbf{Epic-Sounds}\\
& \textbf{(mAP)} & \textbf{(mAP)} & \textbf{(Acc.)} & \textbf{(Acc.)} & \textbf{(Acc.)} & \textbf{(Acc.)}\\
\toprule
AST-S & \textbf{40.32} {\tiny $\pm$ 0.08} & \textbf{29.20} {\tiny $\pm$ 0.11} & \textbf{49.51} {\tiny $\pm$ 0.06} & 39.70 {\tiny $\pm$ 1.83} & 97.38 {\tiny $\pm$ 0.07} & 52.42 {\tiny $\pm$ 0.11}\\
\midrule
\rowcolor{lightgray!25}
\textbf{AuM-S (c)} & {39.68 {\tiny $\pm$ 0.06}} {\tiny \textcolor{nicered}{(-0.64)}} & {28.89 {\tiny $\pm$ 0.20}} {\tiny \textcolor{nicered}{(-0.31)}} & {49.43 {\tiny $\pm$ 0.18}} {\tiny \textcolor{nicered}{(-0.07)}} & {\textbf{40.58} {\tiny $\pm$ 1.11}} {\tiny \textcolor{nicergreen}{(+0.89)}} & {\textbf{97.51} {\tiny $\pm$ 0.08}} {\tiny \textcolor{nicergreen}{(+0.13)}} & {\textbf{52.90} {\tiny $\pm$ 0.40}} {\tiny \textcolor{nicergreen}{(+0.48)}}\\
\bottomrule
\end{tabularx}
\caption{\textbf{Results of Imagenet pretrained initializations of AST and AuM small models across various datasets.} Note that the setup of AuM-S is (c) in Fig. \ref{fig:pipeline} due to the unavailability of the ViM weights for our preferred setup (b).}
\label{tab:smallinit}
\end{table*}

\begin{table}[tp]
\centering
\resizebox{1.0\linewidth}{!}{
\scriptsize
\setlength{\tabcolsep}{2pt} % Adjust the space between columns
\renewcommand{\arraystretch}{1.2} % Adjust the space between rows
\renewcommand{\tabularxcolumn}[1]{m{#1}}
\begin{tabular}{ l ccc ccc }
\toprule
\multirow{2}{*}{\textbf{Model}} & \textbf{VGGSound} & \textbf{VoxCeleb} & \textbf{Speech Comm. V2} & \textbf{Epic-Sounds}\\
& \textbf{(Acc.)} & \textbf{(Acc.)} & \textbf{(Acc.)} & \textbf{(Acc.)}\\
\toprule
AST-B/16 & 44.17 {\tiny $\pm$ 0.14} & \textbf{46.25} {\tiny $\pm$ 1.08} & 90.37 {\tiny $\pm$ 0.06} & 46.62 {\tiny $\pm$ 0.04} \\
\midrule
\rowcolor{lightgray!25}
\textbf{AuM-B/16} & {\textbf{46.61} {\tiny $\pm$ 0.18}} {\tiny \textcolor{nicergreen}{(+2.44)}} & {40.72 {\tiny $\pm$ 1.11}} {\tiny \textcolor{nicered}{(-5.53)}} & {\textbf{94.78} {\tiny $\pm$ 0.04}} {\tiny \textcolor{nicergreen}{(+4.41)}} & {\textbf{48.18} {\tiny $\pm$ 0.13}} {\tiny \textcolor{nicergreen}{(+1.57)}}\\
\bottomrule
\end{tabular}
}
\caption{\textbf{Results of Audioset pretrained initializations of AST and AuM base models across various datasets.}}
\label{tab:baseinit}
\end{table}
\subsection{Ablation Study on Design Choices}
We conduct a series of experiments to verify our design choices and perform further analysis. We study the following strategies in terms of the direction of SSM modules and conv1Ds:

\noindent\underline{\textbf{AuM}-\textbf{Fo}rwardConv1D-\textbf{Fo}rwardSSM:} This choice, which is the default Mamba block, directly applies the AuM Block with only a forward SSM (refer to Figure~\ref{fig:pipeline} (a)).

\noindent\underline{\textbf{AuM}-\textbf{Fo}rwardConv1D-\textbf{Bi}DirectionalSSM:} This is the design of our final model, which applies an additional backward SSM to the previous design choice (refer to Figure~\ref{fig:pipeline} (b)).

\noindent\underline{\textbf{AuM}-\textbf{Bi}DirectionalConv1D-\textbf{Bi}DirectionalSSM:} In this variant, we add another Conv1D in the backward direction to feed the output of this module to the backward SSM, making each SSM module a separate stream. A similar design is adopted in Vision Mamba (ViM) as the default choice (refer to Figure~\ref{fig:pipeline} (c)).

\noindent Moreover, the position of class tokens is ablated for each variant above. To save computational time and resources, we primarily conduct ablation studies by training our model on AudioSet Balanced (AS-20K) and VGGSound. Results are in Table~\ref{tab:ablation}.

\noindent\textbf{Impact of bidirectional SSM.} To understand the impact of the directions of SSM modules, we analyze the performance of the variants of our model with different directional SSM modules. As the results demonstrate, the bidirectional variants (forward and backward SSM modules together) overall show better performance (especially in the large dataset VGGSound) than the forward-only variant.

\noindent\textbf{Impact of direction of conv1D.} Here, we conduct a controlled experiment between two bidirectional methods: \textbf{AuM}-\textbf{Fo}-\textbf{Bi} and \textbf{AuM}-\textbf{Bi}-\textbf{Bi}, where the only difference is the presence of an additional backward Conv1D. As shown in Table~\ref{tab:ablation}, our design choice, which omits the backward Conv1D, generally yields better performance. We hypothesize that processing a single input sequence (the output from only the forward Conv1D) is more effective and natural for scanning in both forward and backward directions to understand entire context, compared to providing separate inputs to each directional SSM module and scanning in only one direction according to the input.

\noindent\textbf{Impact of the class token position.} Our extensive experiments reveal that positioning the class token in the middle of the sequence is the most suitable choice for our design. However, it is important to note that the position of the class token is a crucial decision, as each variant exhibits a different optimal location for its use, which greatly impacts performance. An additional observation is that a forward-only SSM collapses when the class token is placed at the beginning of the sequence (head class token). This outcome is expected, as the information in the sequence following the class token is not incorporated into the class token.

\subsection{Impact of Pre-Training}
\noindent\textbf{Out-of-domain pre-training.} Initializing audio models with ImageNet pre-trained weights has become popular for audio classification~\cite{gong21b_interspeech, chen2022hts}. Specifically, AST demonstrates a significant performance improvement over training from scratch by utilizing the weights of a Supervised ImageNet pretrained ViT model. As presented in Table 1, our main results exclude models with pretraining (weight initialization) to provide a clear comparison between these two architectures. One might question why such results are not displayed. To the best of our knowledge, no released Vision Mamba Base model weights, comparable to ViT weights for the AST model, are available in the literature, preventing us from conducting this experiment directly. However, we aim to analyze both AuM and AST when initialized with out-of-domain pre-training weights. In this context, we utilize the only available Vision Mamba model, the small-sized ViM-S, to compare AuM-S and AST-S models. Despite the differences in architectural design with Vision Mamba, highlighted in Sections 2 and 3.4, where we note that the \textbf{AuM}-\textbf{Bi}-\textbf{Bi} variant is not the ideal choice for our AuM, the findings presented in Table 3 reveal that both models perform similarly. We believe that with the right weight initialization, our model could outperform AST, just as it does in scenarios without the use of vision domain pretrained weights.

\noindent\textbf{From-scratch audio-only pre-training.} After comparing AST and AuM by initializing them with weights from ImageNet pre-trained vision models, we also explore using AudioSet-trained weights of base models (from Table~\ref{tab:baseNOinit}) as in-domain pre-training to initialize both AuM-B and AST-B. Here, unlike in the previous section, our model uses weights from a model that is architecturally identical to ours. The results, shown in Table~\ref{tab:baseinit}, indicate that in-domain pre-training benefits both models, enhancing their performance. In this setting, AuM outperforms AST, except on the VoxCeleb dataset.

\section{Conclusion}
\label{sec:conclusion}
In this work, we introduce Audio Mamba (AuM), the first architecture for audio classification that is free from self-attention and purely based on state space models (SSM). Our extensive experiments highlight AuM's efficiency in terms of computational and memory use, as well as its competitive performance against the well-established Audio Spectrogram Transformers (AST). Considering its similarity to AST structure regarding patchifying the input spectrogram, adding positional embeddings, and processing the information sequentially but without costly self-attention, it shows great potential to become an alternative generic audio backbone. With the elimination of reliance on costly self-attention and the high efficiency of AuM in processing long sequence inputs, we believe that AuM brings an important contribution to the audio field for future potential applications. The ability to handle lengthy audio is increasingly crucial, especially with the rise of self-supervised multimodal learning~\cite{morgado2020learning, alayrac2020self, akbari2021vatt} and generation that leverages in-the-wild data and Automatic Speech Recognition. Furthermore, AuM could be employed in self-supervised learning setups like Audio Masked Auto Encoders~\cite{huang2022masked, baade2022mae} or multimodal learning tasks such as Audio-Visual pretraining~\cite{gong2022uavm, gong2022contrastive, morgado2020learning} or Contrastive Language-Audio Pretraining~\cite{wu2023large, shih2023speechclip, wu2022wav2clip}.
{
    \small
    \bibliographystyle{ieeenat_fullname}
    \bibliography{main}
}

\clearpage
\setcounter{page}{1}
\maketitlesupplementary
\section{Training Setup}
\label{sec:train_setup}
We train all the models (AuM and AST) of all sizes (Base and Small) across six different datasets by following the training setup shown in Table~\ref{tab:train_setup}. 
\begin{table}[t!]
\centering
\renewcommand{\arraystretch}{1.25}
\begin{tabular}{ccccccc}
\hline
\multicolumn{1}{|c}{\textbf{Setting}} & \multicolumn{1}{|c}{\textbf{Audioset}} & \multicolumn{1}{|c}{\textbf{AS-20K}} & \multicolumn{1}{|c}{\textbf{VGGSound}} & \multicolumn{1}{|c}{\textbf{VoxCeleb}} & \multicolumn{1}{|c}{\textbf{Speech Comm. V2}} & \multicolumn{1}{|c|}{\textbf{Epic Sounds}} \\ 
\hline
\multicolumn{1}{|c}{Optimizer} & \multicolumn{6}{|c|}{Adam(wd=5e-7,betas=(0.95, 0.999))} \\
\hline
\multicolumn{1}{|c}{Patch Size / Stride} & \multicolumn{6}{|c|}{16 x 16 / (16, 16)} \\
\hline
\multicolumn{1}{|c}{Batch Size} & \multicolumn{6}{|c|}{12} \\
\hline
\multicolumn{1}{|c}{Weighted Average} & \multicolumn{6}{|c|}{No} \\
\hline
\multicolumn{1}{|c}{Ensembling} & \multicolumn{6}{|c|}{No} \\
\hline
\multicolumn{1}{|c}{Loss Function} & \multicolumn{3}{|c}{BCE} & \multicolumn{3}{|c|}{CE} \\
\hline
\multicolumn{1}{|c}{Multilabel} & \multicolumn{2}{|c}{Yes} & \multicolumn{4}{|c|}{No} \\
\hline
\multicolumn{1}{|c}{Balanced Sampling} & \multicolumn{1}{|c}{Yes} & \multicolumn{5}{|c|}{No} \\
\hline
\multicolumn{1}{|c}{Warm-up Duration} & \multicolumn{5}{|c}{1000 steps} & \multicolumn{1}{|c|}{2 Epochs}\\
\hline
\multicolumn{1}{|c}{Spectrogram Size} & \multicolumn{4}{|c}{128x1024} & \multicolumn{1}{|c}{128x128} & \multicolumn{1}{|c|}{128x1024} \\
\hline
\multicolumn{1}{|c}{SpecAug (time / freq.)} & \multicolumn{4}{|c}{48 / 192} & \multicolumn{1}{|c}{48 / 48} & \multicolumn{1}{|c|}{48 / 192} \\
\hline
\multicolumn{1}{|c}{Mixup} & \multicolumn{2}{|c}{0.5} & \multicolumn{2}{|c}{0} & \multicolumn{1}{|c}{0.6} & \multicolumn{1}{|c|}{0.2} \\
\hline
\multicolumn{1}{|c}{Epochs} & \multicolumn{1}{|c}{5} & \multicolumn{1}{|c}{25} & \multicolumn{2}{|c}{20} & \multicolumn{2}{|c|}{30} \\
\hline
\multicolumn{1}{|c}{LR Sched. Type} & \multicolumn{5}{|c}{MultiStepLR(start / step / decay)} & \multicolumn{1}{|c|}{\multirow{2}{*}{LambdaLR*}} \\
\cline{0-5}
\multicolumn{1}{|c}{LR Sched. Params} & \multicolumn{1}{|c}{2 / 1 / 0.5} & \multicolumn{1}{|c}{10 / 5 / 0.5} & \multicolumn{2}{|c}{5 / 2 / 0.75} & \multicolumn{1}{|c}{5 / 1 / 0.85} & \multicolumn{1}{|c|}{}\\
\hline
\multicolumn{1}{|c}{Dataset Mean for Norm.} & \multicolumn{2}{|c}{-4.268} & \multicolumn{1}{|c}{5.077} & \multicolumn{1}{|c}{-3.761} & \multicolumn{1}{|c}{-6.846} & \multicolumn{1}{|c|}{\multirow{2}{*}{No}}\\
\cline{0-5}
\multicolumn{1}{|c}{Dataset Std. for Norm.} & \multicolumn{2}{|c}{4.569} & \multicolumn{1}{|c}{4.453} & \multicolumn{1}{|c}{4.201} & \multicolumn{1}{|c}{5.565} & \multicolumn{1}{|c|}{}\\
\hline
\multicolumn{1}{|c}{Base LR} & \multicolumn{1}{|c}{1e-5} & \multicolumn{1}{|c}{5e-5} & \multicolumn{2}{|c}{1e-5} & \multicolumn{1}{|c}{2.5e-4} & \multicolumn{1}{|c|}{1e-5} \\
\hline
\vspace{0mm} % helps separating the captionf from the table
\end{tabular}
\makebox[\textwidth]{
\parbox{\textwidth}{
\caption{Training setup comparison across different datasets. Here, "*" indicates that we follow the official learning rate scheduler presented in the Epic Sounds paper.}
\label{tab:train_setup}
}}
\end{table}

\end{document}